# Fractal large-scale structure from a stochastic scaling law model

Salvatore Capozziello[(1)] and Scott Funkhouser[(2)]

[(1)]Dipartimento di Scienze Fisiche, Università di Napoli "Federico II" and INFN Sez. di Napoli, Compl. Univ. Monte S. Angelo, Ed. N., Via Cinthia, I-80126 Napoli, Italy
[(2)]National Oceanic and Atmospheric Administration, 2234 South Hobson Ave., Charleston, SC, 29405-2413, USA

ABSTRACT

A stochastic model relating the parameters of astrophysical structures to the parameters of their granular components is applied to the formation of hierarchical, large-scale structures from galaxies assumed as point-like objects. If the density profile of galaxies on a given scale is described by a power law then the stochastic model leads naturally to a mass function that is proportional to the square of the distance from an occupied point, which corresponds to a two-point correlation function that is inversely proportional to the distance. This result is consistent with observations indicating that galaxies are, on the largest scales, characterized by a fractal distribution with a dimension of order 2 and well-fit with transition to homogeneity at cosmological scales.

## *1. Introduction*

The search for signatures of fractal self-similarity in the large-scale distribution of galaxies was born with the discovery of mathematical fractals such as the Mandelbrot set [1]. Models of a hierarchical, fractal universe are evocative of the galactic hierarchy imagined, for instance, by Charlier [2]. The possibilities of fractal distributions on the largest scales of structure are relevant to modern cosmology for a variety of reasons. If the preponderance of the mass of the universe is distributed in the manner of a fractal on arbitrarily large scales then the cosmological principle, which stipulates that the universe is homogeneous on the largest scales, may require modification [3]. It is important to note, however, that a fractal distribution of galaxies is not necessarily inconsistent with the cosmological principle if there exists a substantial quantity of dark matter [4]. The prevalence of fractal scaling behaviors in the clustering of galaxies may be significant also because it would intimate, perhaps, some new physical mechanism that could be capable of generating fractality at large-scale structure (LSS) such as modified gravity theories where characteristic interaction lengths naturally come out [5].

Let there be some large number of point-like bodies arranged, to some certain extent, in the manner of a fractal. One of the predominant signatures of fractality in such an arrangement is the behavior of the associated mass function $M(r)$, which gives the average, total mass of the bodies contained within a sphere of radius $r$ centered on a given body. In a fractal distribution of point-like bodies there exists at least one structural scale on which $M(r)$ behaves as

$$M(r) \propto r^D, \qquad (1)$$

where $D$ is the fractal dimension. Associated with the mass function (1) is an auto-correlation function $\xi(r)$ of the form

$$\xi(r) \propto r^{-\gamma}, \qquad (2)$$

where $\gamma=3-D$ [6]. The largest $r$ distance for which (1) and (2) are valid is usually stated in terms of so-called correlation length $l$, which is defined by $\xi(l)=1$. There must exist also some lower cut-off, here called $\delta$, for any scale $i$. It is convenient to express (2) in terms of $l$ as

$$\xi(r) \sim \left(\frac{r}{l}\right)^{-\gamma}. \tag{2b}$$

While any given fractal structure may feature other scaling behaviors, (1) and (2) are sufficient for the present analysis.

Modern observations indicate that the distribution of galaxies throughout the observable universe is, to a certain extent, hierarchical and self-similar. The degree to which the arrangement of galaxies is consistent with fractal scaling behaviors may be ascertained by examining the associated auto-correlation function $\xi_g(r)$, which is available from to a variety of detailed galactic surveys. On relatively small scales, $\xi_g(r)$ behaves as a simple power law of the form (2b), with $\gamma \approx 1.8$ and $l \approx 5h^{-1}$Mpc, where $h$ is the Hubble parameter in units of 100km/s/Mpc [6]. Such a distribution corresponds to a fractal mass function of dimension $D = 3 - \gamma \approx 1.2$. On much larger scales, however, the character of the galactic distribution is apparently different. For $r$ as large as perhaps $l \approx 100h^{-1}$Mpc, the two-point correlation function associated with the clustering of galaxies is of the form (2b) with $\gamma \approx 1$ and $D \approx 2$ (See refs. [7] – [14]).

## *2. Fractal structure from the Fluctuative Scaling Law Model*

While the fractal character of large-scale structures is evident empirically, there is no prevailing model of structure formation from which a fractal distribution emerges naturally. The purpose of this paper is to demonstrate that a phenomenological model, based on the stochastic fluctuations of the granular components of astronomical bodies, predicts the formation of LSS in the manner of a fractal hierarchy with dimension of the order 2. Let a large, gravitational system of mass $m_i$ consist primarily of a very large number $N_{ji} \sim m_i/m_j$ of smaller, roughly identical bodies, each having a mass $m_j$. The components $j$ must be point-like with respect to the aggregate system $i$, and may represent either some smaller astronomical body or a particle. The central hypothesis of such a stochastic model is that the characteristic relaxation time $t_i$ of the aggregate body $i$ should be related to the characteristic relaxation time $t_j$ of each component $j$ in a statistical manner according to

$$t_i \sim N_{ji}^{1/2} t_j. \tag{3}$$

The simple, statistical hypothesis in (3) leads to a number of additional scaling relationships. It follows from (3) that the characteristic radius $R_i$ of the aggregate body $i$ should be scaled to the radius $R_j$ of each granular component $j$ according to

$$R_i \sim N_{ji}^{1/2} R_j. \tag{4}$$

The characteristic action $A_i$ of the system $i$ should be scaled to the characteristic action $A_j$ of each component $j$ according to

$$A_i \sim N_{ji}^{3/2} A_j. \tag{5}$$

(See References [15] – [18] and in particular [19]).

Let each body $j$, which represents a granular component of $i$, be also a gravitational, astronomical body that consists primarily of some large number $N_{kj}$ of components $k$. According to this *Fluctuative Stochastic Law Model* (FSLM), characteristic parameters of $j$ and $k$ should be related by (3) – (5) in the same manner in which the parameters of *i* and *j* are related. Furthermore, the body *i* may be treated effectively as an aggregate of components *k* as well as an aggregate of components *j*, and (3) – (5) should relate therefore the parameters of *i* and *k* in the same manner in which *i* and *j* are related. Self-similar hierarchies thus naturally emerge.

The FSLM is, in principle, valid for any gravitational system, and is scale-invariant. The LSS of the observable universe may be treated effectively as aggregates of large numbers of galaxies. According to the FSLM, the characteristic parameters of the LSS should emerge from the stochastic fluctuations of the galaxies. The characteristic mass $m_g$ of a typical large galaxy is empirically of order near $10^{42}$kg, and the characteristic galactic radius $R_g$ is of the order $10^{20}$m. Let us consider some level of hierarchical structure *i* that is populated by bodies that are aggregates of large number $N_{gi}$ of galaxies. If the model is valid, then it follows from (4) that the characteristic radius $R_i$ of each of the aggregate bodies should be related to $R_g$ according to

$$R_i \sim N_{gi}^{1/2} R_g, \tag{6}$$

and thus

$$N_{gi} \sim \frac{R_i^2}{R_g^2}, \tag{7}$$

where $N_{gi} \sim m_i/m_g$.

Let the average number-density $n_g(r)$ of galaxies at some distance *r* relative to the center of a galaxy be given, on any structural scale *i*, by

$$n_g(r) \sim A r^{-a}, \tag{8}$$

for $\delta_i < r < R_i$ where *A* and *a* are positive, real constants and $\delta_i$ is some cut-off that is much smaller than $R_i$. The presumption that the number-density is well described by a spherically symmetric power-law is sufficiently general and model-independent [5,6]. The average number $N_g(r)$ of galaxies contained within a sphere of radius *r* is given by

$$N_g(r) \approx 4\pi \int_{\delta_i}^{r} n_g(\bar{R}) R^2 dR \sim B r^{3-a}, \tag{9}$$

for $a \neq 3$, where $B = 4\pi A/(3-a)$. For $r = R_i$ the number $N_g(r)$ must be equal to $N_{gi}$. It follows therefore from (7) and (9) that

$$B R_i^{3-a} \sim \frac{R_i^2}{R_g^2}, \tag{10}$$

and thus *a*=1 and $B \sim R_g^{-2}$. The average total mass $M_g(r)$ of galaxies contained within a sphere of radius *r* is $m_g N(r)$. From (9) and *a*=1, we have that, for some level *i* of structure and for $\delta_i < r < R_i$,

$$M_g(r) \sim \frac{m_g}{R_g^2} r^2. \tag{11}$$

Note that $(R_i/R_g)^2 \sim N_{gi}$, and (11) thus gives $M_g(R_i) \sim m_g N_{gi}$, which is the desired result, for any scale *i*.

The mass function in (11) has the same form as (1) and characterizes a fractal of dimension 2. A mass function of the form (1) implies a two-point correlation function of the form (2) for $\gamma = 3 - D = 1$, where the correlation length *l* is, for each scale *i*, of the order $R_i$. Since the mass function of galaxies in large-scale structures is empirically given by

(1) with $D \approx 2$, and the associated two-point correlation function is empirically given by (2) with $\gamma \approx 1$, the FSLM leads to a basic requirement concerning the arrangement of large-scale structures that is consistent with observations. The only external premise required for the present result is that the density profile of the distribution of galaxies on a given level of structure is characterized by a simple power law.

The stochastic model is inherently self-similar, and the result in (11) should apply not only to some aggregate of large numbers of galaxies, but also to the structures formed from large numbers of those aggregate bodies. Consider, for instance, structures formed from large numbers of clusters of galaxies, where each cluster has a mass of order near $m_c$ and a radius near $R_c$. Suppose that clusters of galaxies form hierarchical structures whose parameters emerge from the stochastic fluctuations of individual clusters, which may be treated as granular, according to the present model. It follows from (4) that the characteristic radius $R_i$ of some hierarchical structure $i$ formed from clusters should be related to $R_c$ according to

$$R_i \sim N_{ci}^{1/2} R_c , \qquad (12)$$

which is analogous to (6). Since the analysis leading from (6) to (11) is not sensitive to the nature of the granular component, the same analysis should apply to the distribution of clusters, and thus

$$M_c(r) \sim \frac{m_c}{R_c^2} r^2 , \qquad (13)$$

where $M_c(r)$ is the total mass of clusters contained within a sphere of radius $r$, centered on a given cluster or aggregate of clusters. This result is consistent with analyses indicating that the distribution of clusters on the largest scales is characterized by a fractal mass function with dimension 2 [9].

## 3. Conclusions

The observed LSS surveys seem to show a fractal distribution of galaxies and galaxy clusters characterized by fractal dimension of order 2. Considering a stochastic model relating the parameters of astrophysical structures to the parameters of their granular components, the FSLM, the formation of hierarchical LSS is naturally achieved . In particular, if the density profile of galaxies on a given scale is described by a power law, such a stochastic model naturally leads to a mass function that is proportional to the square of the distance from an occupied point, which corresponds to a two-point correlation function that is inversely proportional to the distance. Besides, such a distribution well-fit with transition to homogeneity since the correlation disappears at cosmological scales [6] and FSLM naturally reproduce the cosmological constant according to the ΛCDM model (see [18] for details).